\documentclass[aps,pre,epsfig,showpacs,amsfonts,superscriptaddress,byrevtex]{revtex4}

\usepackage{epsfig}
\usepackage{graphicx}
\begin{document}
%--------------------------------------------------------------------------------------
\title{Complex noise in diffusion-limited reactions of replicating and competing species}
%--------------------------------------------------------------------------------------
\author{David Hochberg}
\email{hochberg@laeff.esa.es} \homepage[]{http://www.cab.inta.es}
\affiliation{Centro de Astrobiolog\'{\i}a (CSIC-INTA), Ctra.
Ajalvir Km. 4, 28850 Torrej\'{o}n de Ardoz, Madrid, Spain}
\author{M.-P. Zorzano}
\email{zorzanomm@inta.es} \affiliation{Centro de
Astrobiolog\'{\i}a (CSIC-INTA), Ctra. Ajalvir Km. 4, 28850
Torrej\'{o}n de Ardoz, Madrid, Spain}
\author{Federico Mor\'an}
\email{fmoran@bio.ucm.es} \affiliation{Centro de
Astrobiolog\'{\i}a (CSIC-INTA), Ctra. Ajalvir Km. 4, 28850
Torrej\'{o}n de Ardoz, Madrid, Spain}
\affiliation{Departamento de
Bioqu\'{\i}mica y Biolog\'{\i}a
Molecular, Facultad de Ciencias Qu\'{\i}micas \\
Universidad Complutense de Madrid, Spain}
%----------------------------------------------------------------------

\date{\today}
\begin{abstract}
We derive exact Langevin-type equations governing quasispecies
dynamics. The inherent multiplicative noise has both real and
imaginary parts. The numerical simulation of the underlying
\textit{complex} stochastic partial differential equations is
carried out employing the Cholesky decomposition for the noise
covariance matrix. This noise produces unavoidable spatio-temporal
density fluctuations about the mean field value. In two dimensions,
the fluctuations are suppressed only when the diffusion time scale
is much smaller than the amplification time scale for the master
species.
\end{abstract}


\maketitle

%----------------------------------------------------------------
\section{\label{sec:intro}Introduction}
%---------------------------------------------------------------

Deterministic descriptions of reacting and diffusing chemical and
molecular species fail to account for the system's internal
fluctuations. Nevertheless, it is known that if the spatial
dimensionality $d$ of the system is smaller than a certain upper
critical dimension $d_c$, these intrinsic fluctuations can play a
crucial role in the asymptotic late time behavior of decay rates
(anomalous kinetics) and the results obtained from the mean field
equations are not correct \cite{KangRedner}. Even far from the
asymptotic regimes, these fluctuations can also control the dynamics
on local spatial and temporal scales \cite{ZHM}. The mean field
result is only valid in the idealized infinite diffusion limit,
because the reactions themselves induce local microscopic density
fluctuations that must be taken into account in the underlying
nonlinear dynamics.

The proper inclusion of the effects of microscopic density
fluctuations in reaction-diffusion systems can be carried out once
the microscopic kinetic equations are specified. With the reaction
scheme in hand, one can derive the corresponding continuous-time
master equation, then represent this stochastic process by
second-quantized bosonic operators and in the final step, pass to a
path integral to map the system onto a continuum stochastic field
theory \cite{Doi,Peliti}. This technique has opened up the way for
employing powerful field-theoretic renormalization group (RG)
methods for studying fluctuations in a number of simple
reaction-diffusion problems \cite{THVL}. Moreover, effective
Langevin-type equations can be deduced from these field-theoretic
actions, in which the noise is made manifest and is specified
precisely. Langevin-type equations are ideally suited for
investigating problems in stability and pattern formation, and lend
themselves for the direct numerical solution of the dynamics.
However, most of the studies devoted to fluctuations in
reaction-diffusion systems are based on applying RG methods to the
field-theoretic actions, with little attention being payed to the
analytical or numerical study of the associated effective Langevin
equations. This is most likely due to the fact that the noise in
this latter representation is often imaginary or even complex, a
feature, which at first glance, may be somewhat surprising
\cite{HT}, and has presented a challenging problem for numerical
simulation.

The purpose of this paper is to confront this imaginary/complex
noise issue at face value. We propose and test out an algorithm for
numerically integrating complex noise in multi-component
reaction-diffusion equations \cite{DFHK}. The model we treat can be
considered as a single quasispecies with error tail, which is
analytically and computationally amenable. The current great revival
of interest in quasispecies dynamics owes to the fact that viral
population dynamics is known to be described by quasispecies
\cite{Domingo}. The major part of this work has been devoted to the
analysis of their dynamics under spatially homogeneous conditions
\cite{ES,Domingo,BE1,BE2,NAMM}. More recently, the importance of
diffusive forces has been recognized and taken into account
\cite{CN,CB}, but rather less attention has been payed to the
presence of the unavoidable internal density fluctuations \cite{HZM}
that are necessarily present in all realistic incompletely mixed
diffusing systems of reacting agents \cite{Epstein}. In view of the
above considerations, it is important to understand how internal
fluctuations affect the evolution of replicator dynamics, and in
what ways do the deterministic and stochastic effects compete. The
specific model treated here maps exactly to a set of Langevin
equations. The advantage for the numerical simulation is that the
model has few fields, the corresponding noise terms are known
exactly, and no approximation is required.

In Section \ref {sec:master} we introduce the specific reaction
scheme. Following a well-established procedure \cite{THVL}, we
derive a field theoretic description of these reactions by means of
the Doi-Peliti formalism \cite{Doi, Peliti}. We obtain the continuum
action, and from this derive an equivalent and exact Langevin
equation description of this quasispecies model. The advantage of
this is that the noise properties are specified automatically and
indicate how the naive mean field reaction-diffusion equations must
be modified to take into account properly the (unavoidable) internal
density fluctuations. The ensuing noise is \textit{complex} and
multiplicative, and in magnitude is controlled by the competition
between the replicator amplification and diffusion. Their numerical
solution is rendered possible by employing the Cholesky
decomposition for the associated noise covariance matrix, as we
describe in detail below in Section \ref{sec:langevin}. In Sec
\ref{sec:numsim} we present results of the numerical simulations of
the complex Langevin equations derived in Sec {\ref{sec:langevin}.
Conclusions are drawn in Sec \ref{sec:disc}.

%---------------------------------------------------------------------
\section{\label{sec:master}From the reaction scheme to the stochastic PDEs}
%---------------------------------------------------------------------

%----------------------------------------------------------------
\subsection{\label{sec:model}The Model}
%---------------------------------------------------------------

We consider a simple replicator model with error introduced via
the faulty self-replication into a mutant species. The mutant
species or, error-tail, undergoes non-catalyzed self-reproduction,
but has no effect on the main species. The system is closed, only
energy can be exchanged with the surroundings, where activated
monomers react to build up self-replicative units. These energy
rich monomers are regenerated from the by-product of the reactions
by means of a recycle mechanism (driven by an external source of
photons--sunlight) maintaining the system out of equilibrium. The
closure of the system directly imposes a selection pressure on the
population. In what follows, ${M}^*,I,I_e$ denote the
concentrations of the activated energy rich monomers, the
replicators and the mutant copies, respectively. The kinetic
constants are introduced in the reaction steps as follows:

Accurate noncatalytic replication with rate $A$:

\begin{equation}\label{ancatrep}
\textrm{M}^* + \textrm{I} \stackrel{AQ}{\longrightarrow}
2\textrm{I}.
\end{equation}

Error noncatalytic replication:

\begin{equation}\label{encatrep}
\textrm{M}^* + \textrm{I} \stackrel{A(1-Q)}{\longrightarrow}
\textrm{I} + \textrm{I}_e.
\end{equation}

Error-species replication with rate $A_e$:

\begin{equation}\label{erep}
\textrm{M}^* + \textrm{I}_e \stackrel{A_e}{\longrightarrow}
2\textrm{I}_e.
\end{equation}

Species degradation and subsequent monomer reactivation with rates
$r,r_e$:
\begin{eqnarray}\label{degradI}
\textrm{I} \stackrel{r}{\longrightarrow} \textrm{M}^*, \\
\label{degradIe} \textrm{I}_e \stackrel{r_e}{\longrightarrow}
\textrm{M}^*.
\end{eqnarray}

The quality factor $Q \in [0,1]$. In order to keep the following
development mathematically manageable, we will assume that the
monomer reactivation step proceeds sufficiently rapidly so that we
can in effect, regard the decay of $I$ and $I_e$ plus the subsequent
reactivation $\textrm{M} \stackrel{energy}{\longrightarrow}
\textrm{M}^*$ as occurring in \emph{one} single step as indicated in
Eq.(\ref{degradI},\ref{degradIe}). If we suppose the system is being
bathed continuously by an external energy source, the monomer
reactivation is occurring continuously, and this should be a
reasonable approximation.  To complete the specification of the
model, we will include spatial diffusion. We allow the $M^*,I$ and
$I_e$ particles to diffuse with constants $D_s,D_I$ and $D_e$,
respectively.  Diffusion is incorporated at the outset in the master
equation. The constraint of constant total particle number is
\textit{automatically} satisfied by the continuous chemical fields
in the mean-field limit, as we demonstrate below. Most importantly,
this constraint provides a selection pressure on the quasispecies.
In the following, we keep the dependence on the model parameters
general, though later on we will choose $D_s >> D_I=D_e$ and
$r=r_e$.

%----------------------------------------------------------------
\subsection{\label{sec:hamiltonian}Mapping to bosonic field theory}
%---------------------------------------------------------------

This chemical master equation for the model
Eqs.(\ref{ancatrep},\ref{encatrep},\ref{erep},\ref{degradI},\ref{degradIe})
can be mapped to a second-quantized description following a
procedure developed by Doi \cite{Doi}. Briefly, we introduce
annihilation and creation operators $a$ and $a^\dag$ for $M^*$, $b$
and $b^\dag$ for $I$ and $c$ and $c^\dag$ for $I_e$ at each lattice
site, with the commutation relations $[a_i,a_j^{\dag}] =
\delta_{ij}$, $[b_i,b_j^{\dag}] = \delta_{ij}$ and $[c_i,c_j^{\dag}]
= \delta_{ij}$. The vacuum state (corresponding to the configuration
containing zero particles) satisfies $a_i|0\rangle = b_i|0\rangle =
c_i|0\rangle = 0$. We then define the time-dependent state vector
\begin{equation}\label{wavefunction}
|\Psi(t)\rangle = \sum_{\{k\},\{m\},\{n\}}P(\{k\},\{m\},\{n\},t)
\prod_i( {a}_i^\dag)^{k_i}( {b}_i^\dag)^{m_i}(
{c}_i^\dag)^{n_i}|0\rangle,
\end{equation}
where $P(\{k\},\{m\},\{n\},t)$ is the probability distribution to
find $k,m,n$ particles of type $M^*,I,I_e$, respectively, at each
site.
The master equation can then be written as a Schr\"{o}dinger-like
equation
\begin{equation}\label{schrodinger}
-\frac{\partial |\Psi(t)\rangle}{\partial t} =
 {H}|\Psi(t)\rangle,
\end{equation}
where the lattice hamiltonian or time-evolution operator is a
function of $a,a^\dag,b,b^\dag,c,c^\dag$ and is given by
\begin{eqnarray}\label{hamiltonian}
 {H} &=& \frac{D_s}{l^2}\sum_{(i,j)}( {a_i}^\dag -
 {a_j}^\dag)( {a_i} -  {a_j} ) +
\frac{D_I}{l^2}\sum_{(i,j)}( {b_i}^\dag -
 {b_j}^\dag)( {b_i} -  {b_j} ) +
 \frac{D_e}{l^2}\sum_{(i,j)}( {c_i}^\dag -
 {c_j}^\dag)( {c_i} -  {c_j} ) \nonumber \\
&-& AQ \sum_i \left[a_i b^{\dag}_i b^{\dag}_i b_i - a^{\dag}_i a_i
b^{\dag}_i b_i \right] - A(1-Q)\sum_i \left[a_i b^{\dag}_i b_i
c_i^{\dag} - a^{\dag}_i a_i b^{\dag}_i b_i \right]\nonumber \\
&-& A_e\sum_i \left[a_i c^{\dag}_i c^{\dag}_i c_i  - a^{\dag}_i
a_i c^{\dag}_i c_i\right]-r \sum_i \left[ a_i^{\dag}b_i -
b_i^{\dag}b_i \right] -r_e \sum_i \left[ a_i^{\dag}c_i -
c_i^{\dag}c_i \right].
\end{eqnarray}
Now take the continuum limit ($l \rightarrow 0$), and obtain a
representation as a path integral \cite{Peliti} over continuous
fields $a(x,t),a^*(x,t),b(x,t),b^*(x,t),c(x,t)$ and $c^*(x,t)$ with
a weight $\exp\big(-S[a,a^*,b,b^*,c,c^*] \big)$, whose action $S$ is
given by
\begin{eqnarray}\label{acion}
S &=& \int dt d^dx \,\left[a^*\partial_t a + D_s\nabla a^* \nabla
a + b^* \partial_t b + D_I\nabla b^* \nabla b + c^*
\partial_t c + D_e \nabla c^* \nabla c -AQ\big(a{b^*}^2 b -a^*ab^*b\big)
\right. \nonumber \\
& & \left. - A(1-Q)\big(ab^*bc^*-a^*ab^*b\big) - A_e\big(a{c^*}^2
c -a^*ac^*c\big) -r\big(a^*b-b^*b\big) -r_e\big(a^*c-c^*c\big)
\right].
\end{eqnarray}
The stationarity conditions (the ``classical field equations")
$\delta S/\delta a = \delta S/\delta b = \delta S/\delta c = 0$ and
$\delta S/\delta a^* = \delta S/\delta b^* = \delta S/\delta c^* =
0$ yield, respectively, $a^* = b^* = c^* = 1$ and the usual
\textit{mean-field} rate equations
\begin{eqnarray}\label{mfa}
\partial_t a &=& D_s\nabla^2 a -Aab-A_eac + rb + r_ec \\
\label{mfb}
\partial_t b &=& D_I\nabla^2 b + AQab - rb \\
\label{mfc}
\partial_t c &=& D_e\nabla^2 c + A(1-Q)ab + A_eac - r_ec .
\end{eqnarray}
We emphasize that these equations represent the mean field
approximation wherein all fluctuations are simply ignored. The
\emph{exact} and correct dynamical equations are fully stochastic,
and we derive them below. Note in the mean field limit, the total
particle number $N$ is automatically conserved: adding up Eq.
(\ref{mfa},\ref{mfb},\ref{mfc}) allows us to prove that for a closed
reaction system (in a bounded and closed reaction domain)
\begin{equation}\label{conserved}
\frac{ d N}{dt} =  \frac{d}{d t}\int d^dx \Big(a(x,t) + b(x,t) +
c(x,t) \Big) = 0,
\end{equation}
so that $N$ is a constant.

%----------------------------------------------------------------
\subsection{\label{sec:Langevin} Equivalent Langevin equation description}
%---------------------------------------------------------------

Here, we go beyond the mean-field approximation Eq.
(\ref{mfa},\ref{mfb},\ref{mfc}) and obtain the exact stochastic
partial differential equations that govern the quasispecies
dynamics. To do so, shift conjugate fields as follows $a^* = 1 +
\tilde a$, $b^* = 1 + \tilde b$, and $c^* = 1 + \tilde c$, then we
can write the action as follows:
\begin{eqnarray}\label{shiftaction}
S &=& \int dt d^dx \,\left[\tilde a\big(\partial_t a -D_s\nabla^2a
+
A ab + A_eac-rb-r_ec\big) \right. \nonumber \\
&+& \left. \tilde b\Big(\partial_tb - D_I\nabla^2b -AQ ab +
rb\big) + \tilde c\big(\partial_tc - D_e\nabla^2c -A(1-Q)ab -A_e
ac + r_ec
\big)\right. \nonumber \\
&-& \left. AQab\big({\tilde b}^2 - \tilde a\tilde b \big)
-A(1-Q)ab\big(\tilde b \tilde c -  \tilde a \tilde b\big) -A_e ac
\big({\tilde c}^2 - \tilde a \tilde c \big)\right].
\end{eqnarray}
The next step is to introduce Gaussian-distributed noise fields
$\eta_a(x,t),\eta_b(x,t),\eta_c(x,t)$ which will permit us to
integrate over the conjugate fields and obtain the exact and
\textit{equivalent Langevin representation} of the stochastic
dynamics contained in $S$.
Now the part in $-S$ quadratic in the conjugate fields $\tilde a,
\tilde b, \tilde c$ contributes to the exponential weight the
following expression (we suppress writing out the $x,t$ dependence
and the integrals $\int d^dx\,dt$ ; these are understood to be
included in what follows):
\begin{eqnarray}\label{quadconj}
\exp(-S)|_{\rm quadratic} &=& \exp\Big( +AQab {\tilde b}^2 - Aab
\tilde a\tilde b +A(1-Q)ab \tilde b \tilde c +A_e ac \big({\tilde
c}^2 - \tilde a \tilde
c\big) \Big)\nonumber \\
&=& \exp\Big( +\mathbf{S}\cdot\mathbf{V}\cdot\mathbf{S}\Big)
\nonumber \\
&=& \exp\sum_{ij} S_i V_{ij} S_j,
\end{eqnarray}
where the vector $\mathbf{S} = (\tilde a,\tilde b,\tilde c)$ and
the $3\times3$ array $\mathbf{V}$ is given by
\begin{equation}\label{v}
\left(%
\begin{array}{ccc}
  0 & -\frac{1}{2}A ab & -\frac{1}{2}A_e ac \\
  -\frac{1}{2}A ab & +AQ ab & +\frac{1}{2}A(1-Q)ab \\
  -\frac{1}{2}A_e ac & +\frac{1}{2}A(1-Q)ab & +A_e ac \\
\end{array}%
\right).
\end{equation}
Now make use of the Hubbard-Stratanovich transformation:
\begin{equation}\label{HS}
\int \prod_id\eta_i\, \exp\left\{-\frac{1}{4}\sum_{ij}\eta_i\big(
V^{-1}_{ij}\big)\eta_j + \sum_i \eta_i S_i \right\} = {\rm
constant}\times \exp\sum_{ij} S_i V_{ij} S_j,
\end{equation}
to express the righthand side of Eq. (\ref{quadconj}) as an integral
over noise fields ($\eta_i$). The covariance matrix $\mathbf{V}$ is
actually a $3 \times 3$ matrix in field space and is proportional to
space and time delta functions (infinite dimensional continuous
``matrices", etc.) We immediately read off the direct and crossed
noise correlations directly from $V_{ij}$, since
\begin{equation}\label{V}
\mathbf{V} = \left(%
\begin{array}{ccc}
  \langle \eta_a \eta_a \rangle & \langle \eta_a \eta_b \rangle & \langle \eta_a \eta_c \rangle \\
  \langle \eta_b \eta_a \rangle & \langle \eta_b \eta_b \rangle & \langle \eta_b \eta_c \rangle \\
  \langle \eta_c \eta_a \rangle & \langle \eta_c \eta_b \rangle & \langle \eta_c \eta_c \rangle \\
\end{array}%
\right).
\end{equation}
For the final step we use Eq. (\ref{HS}) to replace the right hand
side of Eq. (\ref{quadconj}) in Eq. (\ref{shiftaction}). We can now
integrate exactly over the conjugate fields $\tilde a,\tilde
b,\tilde c$, appearing in the path integral $\int {\cal D}a {\cal
D}{\tilde a} {\cal D}b {\cal D}{\tilde b} {\cal D}c  {\cal D}{\tilde
c} \, e^{-S[a,\tilde a, b, \tilde b, c, \tilde c]}$ which yields a
product of delta-functional constraints which imply the following
set of exact coupled set of Langevin equations:
\begin{eqnarray}\label{Langevina}
\partial_t a &=& D_s\nabla^2 a -Aab-A_eac + rb + r_ec + \eta_a,\\
\label{Langevinb}
\partial_t b &=& D_I\nabla^2 b + AQab - rb + \eta_b,\\
\label{Langevinc}
\partial_t c &=& D_e\nabla^2 c + A(1-Q)ab + A_eac - r_ec + \eta_c,
\end{eqnarray}
with noise correlations (compare (\ref{v}) with (\ref{V}))
\begin{eqnarray}\label{noise1}
\langle \eta_a(x,t) \rangle &=&  \langle \eta_b(x,t) \rangle =
\langle \eta_c(x,t) \rangle = 0  \\\label{noise2} \langle
\eta_a(x,t)\eta_a(x',t') \rangle &=& 0  \\\label{noise3} \langle
\eta_b(x,t)\eta_b(x',t') \rangle &=&
+AQa(x,t)b(x,t)\delta^d(x-x')\delta(t-t')
\\\label{noise4}
\langle \eta_c(x,t)\eta_c(x',t') \rangle &=&
+A_ea(x,t)c(x,t)\delta^d(x-x')\delta(t-t')
\\\label{noise5}
\langle \eta_a(x,t)\eta_b(x',t') \rangle &=&
\textstyle{-\frac{1}{2}}A a(x,t)b(x,t)\delta^d(x-x')\delta(t-t')
\\\label{noise6} \langle \eta_a(x,t)\eta_c(x',t') \rangle &=&
\textstyle{-\frac{1}{2}}A_e a(x,t)c(x,t)\delta^d(x-x')\delta(t-t')
\\\label{noise7} \langle \eta_b(x,t)\eta_c(x',t') \rangle &=&
\textstyle {+\frac{1}{2}} A(1-Q)a(x,t)b(x,t)
\delta^d(x-x')\delta(t-t').
\end{eqnarray}

Note the noise $\eta_a$ has finite cross correlations
Eqs.(\ref{noise5},\ref{noise6}) but \textit{zero} autocorrelation
Eq.(\ref{noise2}). This is already a indicator that the noise cannot
be purely real. In the following Section, we will show that the
pattern of the above correlations is solved by complex noise.

%------------------------------------------------------------------
\section{\label{sec:langevin}Langevin equations with complex noise}
%-----------------------------------------------------------------

It is reasonable to assume that both the replicating and mutant
species diffuse with equal rates $D_{Ie}=D_I=D$ and have equal
degradation rates, i.e. $r_e/r=1$. We thus consider the following
reaction-diffusion system, in a two-dimensional space, subject to
noise and employing the \textit{dimensionless} fields, noises and
model parameters (for the details of the non-dimensionalization of
the stochastic reaction-diffusion system, see Appendix
\ref{sec:appendix}):
\begin{eqnarray}\label{QEN}
\frac{\partial \bar{a}}{\partial \tau}&=&(D_s/D) \hat \nabla ^{2}
\bar{a}  - \bar{a}\bar{b}-\bar{a}\bar{c}+\bar{b}+
\frac{1}{\delta}\bar{c} + \hat{\eta}_a\\ \nonumber \frac{\partial
\bar{b}}{\partial \tau}&=&\hat \nabla ^{2} \bar{b}  +Q
\bar{a}\bar{b}- \bar{b} + \hat{\eta}_b \\ \nonumber \frac{\partial
\bar{c}}{\partial \tau}&=&\hat \nabla ^{2} \bar{c}  + \delta
(1-Q)\bar{a}\bar{b}+ \delta\bar{a}\bar{c} - \bar{c}+ \hat{\eta}_c,
\end{eqnarray}
where $\hat \nabla ^{2}=\frac{\partial ^{2}}{\partial \hat
x^{2}}+\frac{\partial ^{2}}{\partial \hat y^{2}}$ and
$\vec{{\eta}}=(\hat{\eta} _b,\hat{\eta} _c,\hat{\eta} _a)$ is the
noise vector defined above. The initial condition
$\bar{a}_0,\bar{b}_0,\bar{c}_0$ and the ratio of replication rates
$\delta = \frac{A_e}{A} < 1$ obeys the dimensionless constraint for
the closed system $\bar{N}=\int \int
d\hat{x}d\hat{y}(\bar{a}_0+\bar{b}_0+\frac{\bar{c}_0}{\delta})$. In
the two-dimensional case the total number of particles is given by
the ratio $N=\bar{N}/\epsilon$ where $\epsilon=A/D_I$ is the ratio
of the reaction to the diffusion processes (in any dimension $d$, we
have $\epsilon=(r/D_I)^{d/2}A/r$).
%For $d=2$, $N = {\bar N/\epsilon}$.

In the deterministic case and per each single cell $\delta \hat x
\times \delta \hat y$,  this reaction-diffusion system Eq.
(\ref{QEN}) has the following set of homogeneous and static
solutions:
\begin{itemize}
\item{$\bar{b}=\bar{c}=0,\ \bar{a}=\bar{N}$, absorbing solution}
\item{$\bar{b}=0,\ \bar{c}/\delta=\bar{N}-\frac{1}{\delta},\
 \bar{a}=\frac{1}{\delta}$, if $\delta>1/\bar{N}$}
\item{$\bar{c}/\delta=\frac{(\bar{N}Q-1)(1-Q)}{Q(1-\delta)},\
 \bar{b}=\frac{(\bar{N}Q-1)(Q-\delta)}{Q(1-\delta)}$, $\bar{a}=1/Q$
if $Q>\delta$ and $Q>1/\bar{N}$}.
\end{itemize}
For convenience we can reorder the noise vector components
$\vec{{\eta}}=(\hat{\eta} _b,\hat{\eta} _c,\hat{\eta} _a)$ such that
it has the correlation matrix:
\begin{equation}\label{B}
B=<\vec{{\eta}}\vec{{\eta}}'^T>=\left(
\begin{array}{ccc}
\epsilon Q \bar{a} \bar{b} & \frac{1}{2}\epsilon\delta (1-Q)
\bar{a} \bar{b} &  -\frac{1}{2}\epsilon \bar{a} \bar{b} \\
\frac{1}{2}\epsilon\delta (1-Q)\bar{a} \bar{b} & \epsilon \delta^2
\bar{a} \bar{c} &  -\frac{1}{2}\epsilon \delta \bar{a} \bar{c}\\
-\frac{1}{2}\epsilon \bar{a} \bar{b} & -\frac{1}{2}\epsilon \delta
\bar{a} \bar{c} & 0
\end{array}
\right),
\end{equation}
where the zero autocorrelation term is located in the last column
and last row. Notice that $B$ is a symmetric matrix with $\det
B=-\frac{Q}{4}\bar{a}^3 \bar{b} \bar{c} \epsilon ^3 \delta
^2(\bar{b} +\bar{c} )$, i.e. it is negative definite.

For an $M\times M$ symmetric matrix $B$, one can apply the Cholesky
decomposition $B=L L^T$ to extract the square root of the matrix in
the form of a lower triangular matrix $L$ with
$L_{ii}=\sqrt{B_{ii}-\sum_{k=1}^{i-1}L^2_{ik}}$ and
$L_{ji}=\frac{1}{L_{ii}}(B_{ji}-\sum_ {k=1}^{i-1}L_{ik}L_{jk})$ and
$j=i+1,...,M$. This decomposition is used when the symmetric matrix
is positive definite. We have applied this algorithm to our case
with negative definite correlation matrix and we obtain then the
matrix ``square root'' $L$ where some of the terms are manifestly
imaginary:
\begin{equation}
L=\sqrt{\epsilon \bar{a}}\left(
\begin{array}{ccc}
\sqrt{\bar{Qb}} & 0 & 0 \\
\frac{\delta}{2}(1-Q)\sqrt{\bar{b}/Q} &
\frac{\delta \sqrt{4Q\bar{c}-(1-Q)^2 \bar{b}}}{2\sqrt{Q} } & 0 \\
-\frac{1}{2}\sqrt{\bar{b}/Q} &
  \frac{1}{2\sqrt{Q} } \frac{(1-Q)\bar{b}-2 Q\bar{c}}{\sqrt{4Q\bar{c}-(1-Q)^2 \bar{b}}} &
  \sqrt{-1}\sqrt{Q}\frac{\sqrt{ \bar{b}\bar{c}+\bar{c}^2}}{\sqrt{ 4Q\bar{c}-(1-Q)^2 \bar{b}}}
\end{array}
\right).
\end{equation}
We will use this decomposition to relate the noise to a new {\em
real} noise $\vec{\xi}$, a white Gaussian noise with
$<\vec{\xi}\vec{\xi}'^T>=1$ (i.e. with uncorrelated real
components) such that $\vec{\eta}=L\vec{\xi}$ (and thus
$\vec{\eta}^T=\vec{\xi}^TL^T$). We thus are able to write the
internal noise as a linear combination of white noise terms.
Notice that by doing so the condition
$<\vec{\eta}\vec{\eta}'^T>=<L\vec{\xi}\vec{\xi}'^TL^T
>=L<\vec{\xi}\vec{\xi}'^T>L^T=L L^T=B$ is satisfied. This
transformation will allow us to separate the real and imaginary
parts of the noise, a useful feature to have for setting up a
numerical simulation of the system in Eq. (\ref{QEN}).

Given three uncorrelated (real) white Gaussian noises $\xi_{1}$,
$\xi_{2}$ and $\xi_{3}$ (the three components of $\vec{\xi}$) we
recover $\hat{\eta}_{b}$, $\hat{\eta}_{c}$ and $\hat{\eta}_{a}$
with the specified cross-correlations Eqs.
(\ref{dnoise2}-\ref{dnoise7}) by writing:
\begin{equation}
\begin{array}{c}
\hat{\eta_{b}}=\sqrt{\epsilon \bar{a}}\ \sqrt{\bar{Qb}}\,\xi_{1}\\
\hat{\eta_{c}}=\sqrt{\epsilon
\bar{a}}\ \left(\frac{\delta}{2}(1-Q)\sqrt{\bar{b}/Q}\xi_{1}+
\frac{\delta \sqrt{4Q\bar{c}-(1-Q)^2 \bar{b}}}{2\sqrt{Q} }\xi_{2}\right) \\
\hat{\eta_{a}}=\sqrt{\epsilon \bar{a}}\
\left(-\frac{\sqrt{b}}{2\sqrt{Q} }\xi_{1}+\frac{1}{2\sqrt{Q} }
\frac{(1-Q)\bar{b}-2 Q\bar{c}}{\sqrt{4Q\bar{c}-(1-Q)^2
\bar{b}}}\xi_{2} +   \sqrt{-1}\sqrt{Q}\frac{\sqrt{
\bar{b}\bar{c}+\bar{c}^2}}{\sqrt{ 4Q\bar{c}-(1-Q)^2
\bar{b}}}\xi_{3}\right).
\end{array}
\end{equation}
The noise $\hat{\eta_{a}}$ for the nutrient field {\em always} has
an imaginary component and, since this noise is {\em feeding} the
nutrient field reaction-diffusion equation, the nutrient field also
has an imaginary component. Finally since all the equations are
coupled to the nutrient $\bar{a}$, even if the initial condition for
$\bar{a},\bar{b},\bar{c}$ is real, the fields will, in principle,
have both real and imaginary part and thus we have to solve a system
of six partial differential equations, one for each field (${\rm Re}
(\bar{a}),{\rm Im}(\bar{a}),{\rm Re}(\bar{b}), {\rm
Im}(\bar{b}),{\rm Re} (\bar{c}), {\rm Im}(\bar{c})$) with
two-dimensional diffusion and noise. At this juncture, it is
important that we confirm numerically that the complex solutions,
once averaged over the fluctuations, do indeed yield real results,
in accord with the theoretical expectations \cite{Cardy}.

%--------------------------------------------
\section{\label{sec:numsim}Numerical results}
%--------------------------------------------

By inspection of the correlation matrix (\ref{B}) we see that the
internal, unavoidable, reaction noise is multiplicative and its
intensity is proportional to $\epsilon$. The parameter $\epsilon$ is
the ratio of the reaction to the diffusion processes (in $d=2$,
$\epsilon=A/D_I$ and $\epsilon\delta=A_e/D_e$).  The multiplicative
noise is proportional both to $\bar{a}$, and either $\bar{b}$ or a
combination of $\bar{b}$ with $\bar{c}$. If the system stays close
to the mean field result, then $\bar{a}=1/Q$ and $\bar{b}$ and
$\bar{c}$ scale as $(\bar{N}Q-1)$. Finally, the multiplicative noise
will increase with both increasing $\epsilon$ and/or increasing
$\bar{N}$. Only when the diffusion processes dominate and $\epsilon
\rightarrow 0$, does the noise terms vanish, the local details of
reaction are erased, and we expect to recover the homogeneous
solution of the mean-field approximation. Our main interest here is
the limit when the diffusion does not dominate, $\epsilon>0$ and the
noisy multiplicative term has a significant contribution. In this
regime the problem can thus not be analyzed by perturbation theory
and has to be treated numerically.

The numerical simulations of system evolution have been performed
using forward Euler integration of the finite-difference equations
following discretization of space and time in the stochastic partial
differential equations. The spatial mesh consists of a lattice of
$154\times 154$ cells with cell size $\Delta \hat x=\Delta \hat y=
0.35$ and periodic boundary conditions. Noise has been discretized
as well. The system has been numerically integrated up to $\tau=100$
(with time step $\Delta \tau=2.5\times 10^{-4}$). Integrating the
system numerically we confirmed that the imaginary fields were zero
in average as expected, since the stochastic averages $\langle a
\rangle, \langle b\rangle$, and $\langle c \rangle$ correspond to
the physical densities and thus the scaled number of particles was
preserved and remained {\em real}: ${\rm Re}(\bar{N})=\bar{N}$, and
${\rm Im}(\bar{N})=0$ (within computational errors)  \footnote{In
fact, for computational purposes, ${\rm Im}(\bar{N})$ was set as
small as possible. We confirmed that both ${\rm Re}(\bar{N})=\int
\int d\hat{x}d\hat{y}({\rm Re}(\bar{a})+ {\rm
Re}(\bar{b})+\frac{{\rm Re}(\bar{c}_0)}{\delta})=\bar{N}$ and ${\rm
Im}(\bar{N})=\int \int d\hat{x}d\hat{y}({\rm Im}(\bar{a})+ {\rm
Im}(\bar{b})+\frac{{\rm Im}(\bar{c}_0)}{\delta})=\bar{N}\times10
^{-7}$ are conserved during the integrating time and that the
results were independent of ${\rm Im}(\bar{N})$.}.

This method is able to provide the time evolution of the spatial
distribution of each of the fields. As an example we consider the
evolution of the spatially uniform initial condition,
$\bar{a}=1/Q,\bar{b}=\frac{(\bar{N}Q-1)(Q-\delta)}{Q(1-\delta)},\bar{c}/
\delta=\frac{(\bar{N}Q-1)(1-Q)}{Q(1-\delta)}$ (which is the solution
of mean-field problem) for $D_s/D=10$, $Q=0.92,\delta=0.8$ and
$\bar{N}=100$ and noise intensity $\epsilon=1$. In Fig.
\ref{spatial} we show the time evolution of the real part of the
distributions of the master ($\bar b$) and mutant ($\bar c$)
species, which exhibit spatial fluctuations with respect to the
mean-field value (up to $+1\%$ in black and $-1\%$ in white). The
spatial distributions of the master and mutant species are
anti-correlated. After a transient time, the short-range spatial
fluctuations (e.g., at $\tau = 0.5$) evolve into long-range spatial
fluctuations (e.g., at $\tau = 75$) with wider regions of higher
densities of either one or the other species.
\begin{figure}[h]
\begin{center}
\begin{tabular}{ccc}
\includegraphics[width=0.2 \textwidth]{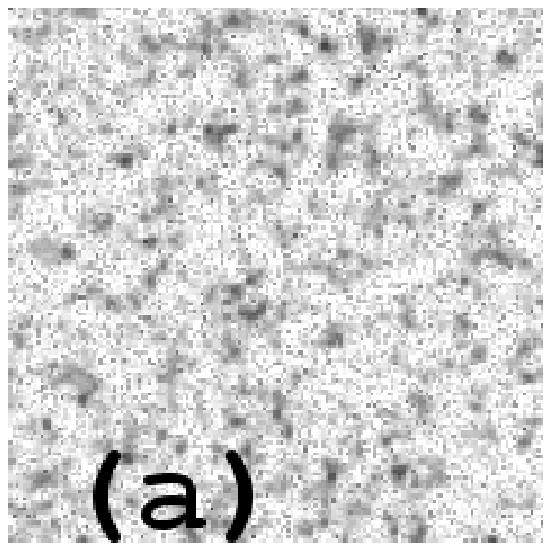}&
\includegraphics[width=0.2 \textwidth]{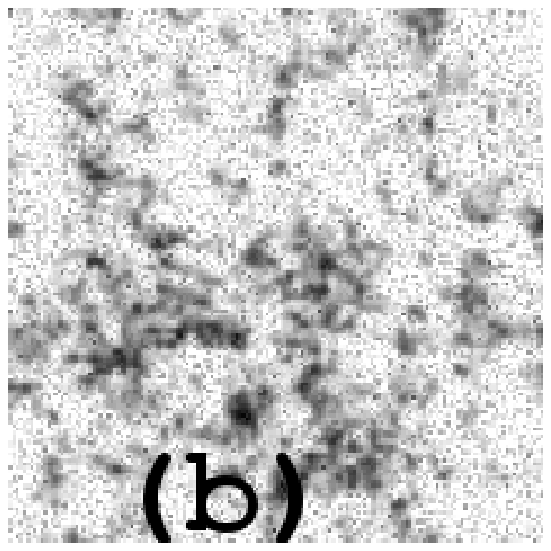}&
\includegraphics[width=0.2 \textwidth]{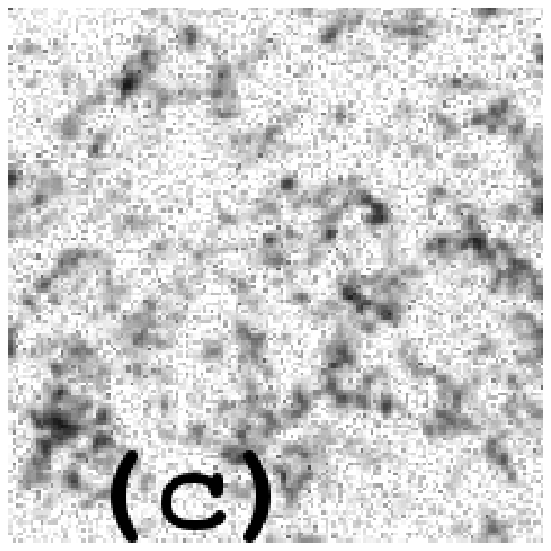}\\
\includegraphics[width=0.2 \textwidth]{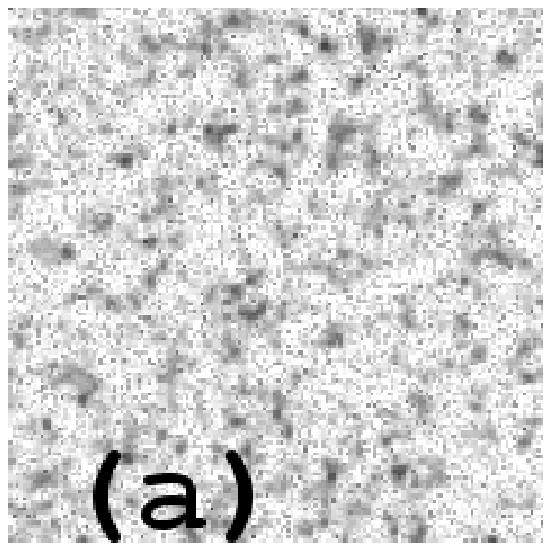}&
\includegraphics[width=0.2 \textwidth]{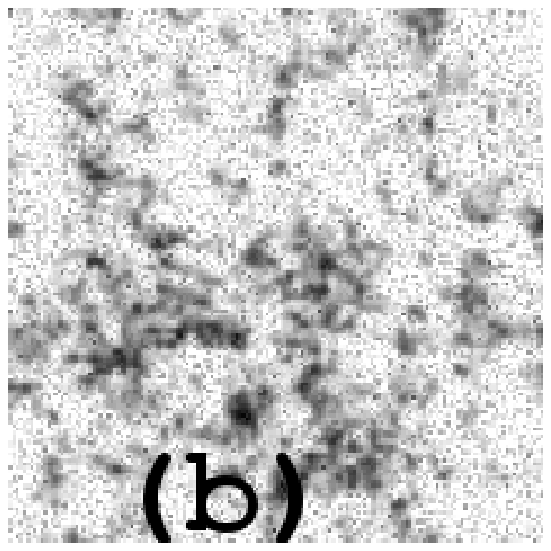}&
\includegraphics[width=0.2\textwidth]{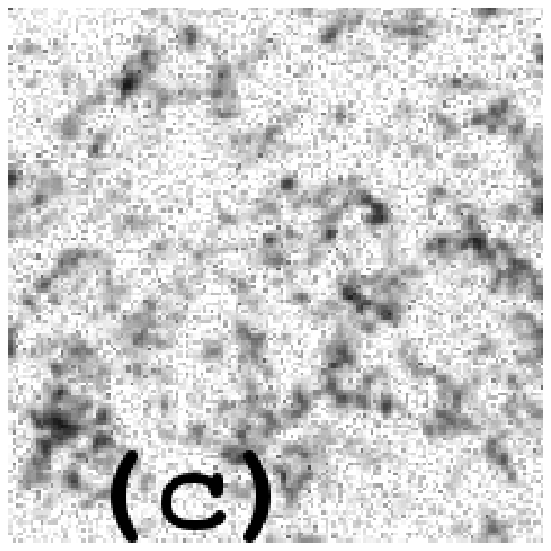}\\
\end{tabular}
\caption{\label{spatial}Temporal sequence of spatial fluctuations in
the real part of the stochastic fields
$\bar{b}(\hat{x},\hat{y},\tau)$ (upper row) and
$\bar{c}(\hat{x},\hat{y},\tau)$ (lower row). The spatial
distributions are shown in (a) at $\tau=0.5$,  in (b) at $\tau=50$
and  in (c) at $\tau=75$.}
\end{center}
\end{figure}
The system converges always to the same {\em spatially averaged}
solution independently of the initial condition and noise intensity.
For example, in Fig. \ref{fig2}-(Left) we show for  $D_s/D=10$,
$Q=0.92,\delta=0.8$ and $\bar{N}=100$, the time evolution of the
{\em spatially averaged} solution ($<\bar{a} (\tau)>= \int \int
d\hat{x} d\hat{y}\ \bar{a}(\hat{x},\hat{y},\tau)$, etc.) for two
noise intensities, $\epsilon=0.1$ and $\epsilon=10^{-2}$, when the
initial condition is set to
$\bar{a}(\hat{x},\hat{y},0)=\bar{b}(\hat{x},\hat{y},0)=\bar{c}(\hat{x},\hat{y},0)/\delta=\bar{N}/3$.
We can see that independently of the noise intensity this averaged
value converges to the solution of the mean-field problem
$\bar{a}=1/Q,\bar{b}=\frac{(\bar{N}Q-1)(Q-\delta)}
{Q(1-\delta)},\bar{c}/\delta=\frac{(\bar{N}Q-1)(1-Q)}{Q(1-\delta)}$.
Higher order moments, such as the mean square value, can be also
extracted from the distribution. For the same study case we the show
the time evolution of the mean square value ($\sigma_{\bar{b}
(\tau)}=\sqrt{<(\bar{b} (\tau)-<\bar{b} (\tau)>)^{2}>}$) for three
noise intensities, $\epsilon=1$, $\epsilon=0.1$ and $\epsilon=0.01$,
Fig. \ref{fig2}-(Right), confirming that only in the limit
$\epsilon\rightarrow 0$ do spatio-temporal fluctuations vanish and
we recover the homogeneous stationary solution of the mean-field
approximation.
\begin{figure}[h]
\begin{center}
\begin{tabular}{cc}
\includegraphics[width=0.4 \textwidth]{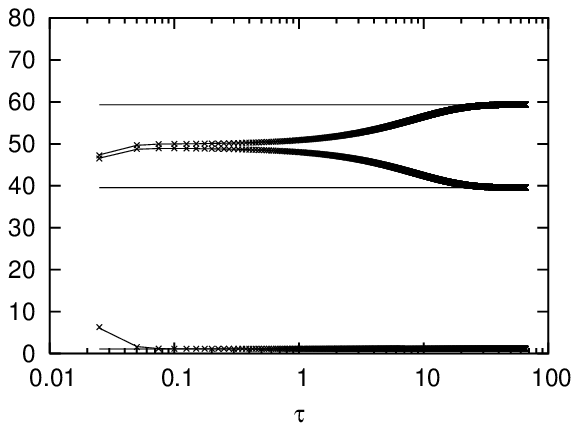}&
\includegraphics[width=0.4 \textwidth]{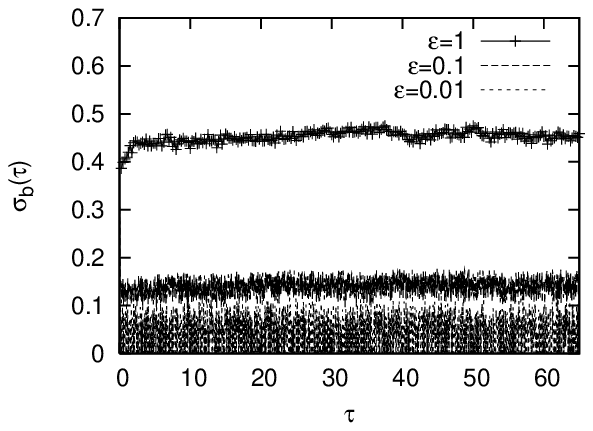}
\end{tabular}
\caption{\label{fig2} (Left) Time evolution of spatially averaged
values (upper curve for $<\bar b>$, middle for $<\bar c>/\delta$ and
lower for $<\bar a>$) for $\bar{N}=100$,  $\epsilon=0.1$ and
$\epsilon=0.01$ (the difference in these two noise levels is not
distinguishable in the figure) when the initial condition was set to
$\bar{a}(\hat{x},\hat{y},0)=\bar{b}(\hat{x},\hat{y},0)=\bar{c}
(\hat{x},\hat{y},0)/\delta=\bar{N}/3$ . The {\em spatially averaged}
solution always converges to the mean-field solution which is marked
by the horizontal lines. (Right) Time evolution of the mean square
value $\sigma_{\bar{b} (\tau)}$ for $\epsilon=1$, $\epsilon=0.1$ and
$\epsilon=0.01$ showing that the spatial fluctuations scale with the
noise intensity, vanishing only in the limit $\epsilon\rightarrow
0$.}
\end{center}
\end{figure}

%----------------------------------------------------------------
\section{\label{sec:disc}Discussion}
%---------------------------------------------------------------

Starting from the microscopic kinetic equations for a single master
species competing with mutants for a limited pool of nutrients, we
have derived a set of stochastic partial differential equations that
exactly describe the complete dynamics of this closed reacting and
diffusing system. The unavoidable fluctuations inherent to the
system give rise to multiplicative noise having both real and
imaginary components. A procedure is presented to solve numerically
this set of partial differential equations. The internal noise
associated with the microscopic details of the reaction produces
unavoidable spatio-temporal density fluctuations around the mean
field value. We estimate the size of these fluctuations. They
strictly vanish, and the mean-field limit is recovered, only when
the diffusion processes are much faster than the rate of master
species amplification. We find that the averages $\langle \bar a
\rangle,\langle \bar b \rangle$ and $\langle \bar c \rangle$ tend to
stationary values (see Fig \ref{fig2}-(Left)  ) as do also the
second moments, or variances, $\sigma_{\bar{a} (\tau)},
\sigma_{\bar{b} (\tau)}$ and $\sigma_{\bar{c} (\tau)}$, see Fig
\ref{fig2}-(Right).

The general purpose algorithm presented here expresses a set of
complex Gaussian noises with defined covariance matrix as a linear
combination of real, white Gaussian noises. This allows for the
numerical generation of this noise, and thus the numerical
integration of Langevin-type equations with complex noise. We have
performed numerical simulations with these dynamical equations to
assess directly the influence of this noise on the evolution of the
stochastic fields associated with a quasispecies, its error tail and
the activated monomer distributions. Apart from the specific
application to the problem analyzed in this paper, we believe this
matrix decomposition will be of great practical use for simulating
other stochastic PDEs with complex noise, a subject which has
received little attention up to now \cite{DFHK}. In this regard it
is interesting to emphasize that the appearance and necessity of
complex and/or imaginary noise in probabilistic descriptions of both
quantum optics and nonlinear chemical reaction systems has been
recognized for some time now and began to be placed on a firm
theoretical footing well over twenty years ago \cite{DrummondGW}.

The simulations are carried out in $d=2$ space dimensions. For this
dimension, the basic amplitude of the noise is controlled by the
dimensionless parameter $\epsilon = A/D_I$, where $A$ denotes the
amplification rate of the quasispecies and $D_I$ its diffusion.
Thus, the noise is controlled, in part, by the competition between
production and diffusion. Diffusion tends to erase or smooth out
local concentration gradients while particle production, which
occurs locally, tends to increase them (i.e., increases the local
density of the species being amplified). In $d=2$ and using Eq.
(\ref{dims}), we can also write $\epsilon = t_D/t_A$, where
$t_D,t_A$ denote the diffusion and species amplification time
scales, respectively. This shows that the noise arises through the
competition between these two time scales. Furthermore, since the
noise is \textit{multiplicative}, its amplitude also depends on the
bilinear product of concentration fields (see, e.g., Eqs.
(\ref{dnoise1}- \ref{dnoise7})) so that whenever monomer and
replicator or monomer and error tail meet at a point and react, this
gives rise to noise at that point, whose strength is directly
proportional to the product of particle concentrations at that point
(equal to the product of particle numbers per unit area). Thus the
complete noise amplitude is modulated by $\epsilon$ \textit{and} the
local bilinear concentrations. So, while diffusion smooths out
inhomogeneities and tends to diminish the internal noise, increasing
the total fixed particle number $N$ (in a bounded domain) leads to
stronger local fluctuations. This is easy to understand since, in
spite of the diffusion, increasing $N$ leads to a ``pile-up" of
reactants at spatial points. There will always be fluctuations about
this approximate homogeneous state, and the statistical deviation
from homogeneity grows with increasing total particle number.

In realistic situations with finite diffusion rates, the system does
not converge to a homogeneous solution with a unique defined value,
but to a state where the densities fluctuate both in space and time
around a mean value. This approach is useful to estimate the
expected deviations of the densities with respect to the mean-field
value when the microscopic reaction details are taken into account.
For systems with higher degree of non-linearity these deviations may
eventually lead the system to new asymptotic states and also induce
the formation of true patterns. We point out that no true spatial
patterns are generated by the underlying reaction diffusion model
studied in this paper. The model as it stands is weakly non-linear,
depending only quadratically on the fields and these second-order
reaction terms are not capable of giving rise to spatial patterns.
The proper inclusion of noise induces random structures. However,
once catalyzed self-replication is included, thereby leading to a
bona-fide \textit{network} of quasispecies, the underlying dynamics
becomes cubically nonlinear, and cubic terms can lead to a variety
of possible spatial patterns \cite{CN} whose evolution and stability
properties in presence of internal noise can be profitably studied
with the methods offered in this paper \cite{HZM}. Finally, we
mention that the methods in \cite{Doi},\cite{Peliti} were used
recently to study the critical behavior of a simple model of
quasispecies in the vicinity of the error-threshold \cite{PSP}.

\begin{acknowledgments}
We thank Carlos Escudero for useful discussions and for
independently working out some preliminary analytic calculations
inspired by our model. M.-P.Z. is supported by a fellowship provided
by INTA for training in astrobiology. The research of D.H. is
supported in part by funds from CSIC and INTA and F.M. is supported
in part by grant BMC2003-06957 from MEC (Spain).
\end{acknowledgments}

%-------------------------------------------------------
\appendix
\section{\label{sec:appendix} Nondimensionalization}
%-------------------------------------------------------

For the purposes of numerical simulation, it is convenient to cast
the system of stochastic partial differential equations Eqs.
(\ref{Langevina}-\ref{Langevinc}) and noise statistics Eqs.
 (\ref{noise1}-\ref{noise7})in terms of dimensionless fields and
parameters. To this end, define basic length and time scales $L$
and $T$, respectively. Then dimensional analysis of any of the
three stochastic equations in Eqs.
(\ref{Langevina}-\ref{Langevinc}) yields
\begin{eqnarray}\label{dims}
&& [a] = [b] = [c] = L^{-d},\qquad [D_I] = [D_e] = [D_s] = L^2/T
\qquad [Q] = 1, \qquad [A] = [A_e] = L^d/T, \\
&& [r] = [r_e] = T^{-1}, \qquad [\eta_a] =[\eta_b] = [\eta_c] =
T^{-1}L^{-d}.
\end{eqnarray}
Define the dimensionless fields:
\begin{equation}\label{dimlessfields}
\bar a = \frac{A}{r}a, \qquad \bar b = \frac{A}{r}b, \qquad \bar c
= \frac{A_e}{r}c,
\end{equation}
and the dimensionless time and spatial coordinates and their
corresponding derivative operators:
\begin{eqnarray}\label{dimlesstime}
&&\tau = rt, \qquad \hat x_j = \Big(\frac{r}{D_I}\Big)^{1/2}\,x_j,\\
\Rightarrow&&  \frac{\partial}{\partial \tau} =
\frac{1}{r}\frac{\partial}{\partial t}, \qquad {\hat \nabla}^2 =
\big(\frac{D_I}{r}\big)\nabla^2.
\end{eqnarray}
We obtain the following dimensionless version of the stochastic
equations listed in Eqs. (\ref{Langevina}-\ref{Langevinc}):
\begin{eqnarray}\label{dimlessLangevina}
\partial_\tau \bar a &=& \Big(\frac{D_s}{D_I}\Big){\hat \nabla}^2 \bar a - {\bar a}{\bar b}
-{\bar a}{\bar c} + {\bar b} + \Big(\frac{r_e A}{r A_e}\Big){\bar c} + {\hat \eta_a},\\
\label{dimlessLangevinb}
\partial_\tau \bar b &=& {\hat \nabla}^2 \bar b + Q{\bar a}{\bar b} - {\bar b} + {\hat \eta}_b,\\
\label{dimlessLangevinc}
\partial_\tau \bar c &=& \Big(\frac{D_e}{D_I}\Big){\hat\nabla}^2 \bar c + \frac{A_e}{A}(1-Q)
{\bar a}{\bar b} + \frac{A_e}{A}{\bar a}{\bar c} -
\frac{r_e}{r}\bar c + {\hat \eta_c},
\end{eqnarray}
where the dimensionless noises are defined by
\begin{equation}\label{dimlessnoise}
{\hat \eta_a} = \frac{A}{r^2}\eta_a, \qquad {\hat \eta_b} =
\frac{A}{r^2}\eta_b, \qquad {\hat \eta_c} = \frac{A_e}{r^2}\eta_c,
\end{equation}
and the dimensionless noise correlations are
\begin{eqnarray}\label{dnoise1}
\langle \hat \eta_a(\hat x,\tau ) \rangle &=&  \langle \hat
\eta_b(\hat x,\tau) \rangle = \langle \hat \eta_c(\hat x,\tau)
\rangle = 0
\\\label{dnoise2} \langle \hat \eta_a(\hat x,\tau)\hat \eta_a(\hat x',\tau') \rangle &=& 0
\\\label{dnoise3} \langle \hat \eta_b(\hat x,\tau)\hat \eta_b(\hat x',\tau') \rangle &=&
+\epsilon\, Q{\bar a(\hat x,\tau)}{\bar b(\hat
x,\tau)}\delta^d(\hat x -\hat x')\delta(\tau-\tau')
\\\label{dnoise4}
\langle \hat\eta_c(\hat x,\tau)\hat\eta_c(\hat x',\tau') \rangle
&=& +\epsilon \delta^2\, {\bar a(\hat x,\tau)}{\bar c(\hat
x,\tau)}\delta^d(\hat x- \hat x')\delta(\tau-\tau')
\\\label{dnoise5}
\langle \hat\eta_a(\hat x,\tau)\hat\eta_b(\hat x',\tau') \rangle
&=& \textstyle{-\frac{1}{2}}\epsilon\, {\bar a(\hat x,\tau)}{\bar
b(\hat x,\tau)}\delta^d(\hat x- \hat x')\delta(\tau-\tau')
\\\label{dnoise6} \langle \hat\eta_a(\hat x,\tau)\hat\eta_c(\hat x',\tau') \rangle &=&
\textstyle{-\frac{1}{2}}\epsilon  \delta\, {\bar a(\hat
x,\tau)}{\bar c(\hat x,\tau)}\delta^d(\hat x- \hat
x')\delta(\tau-\tau')
\\\label{dnoise7} \langle \hat\eta_b(\hat x,\tau)\hat\eta_c(\hat x',\tau') \rangle &=&
\textstyle {+\frac{1}{2}}\epsilon \delta \, (1-Q){\bar a(\hat
x,\tau)}{\bar b(\hat x,\tau)} \delta^d(\hat x- \hat
x')\delta(\tau-\tau').
\end{eqnarray}
In arriving at this, note that
\begin{equation}
[\delta^d(x)] = L^{-d}, \qquad [\delta(t)] = T^{-1},
\end{equation}
whereas the following deltas are dimensionless:
\begin{equation}
[\delta^d(\hat x)] = 1, \qquad [\delta(\tau)] = 1,
\end{equation}
and
\begin{equation}
\delta^d(x) \delta(t) = r
\Big(\frac{r}{D_I}\Big)^{d/2}\delta^d(\hat x)\delta(\tau).
\end{equation}

The basic control parameters are given by
\begin{equation}
\epsilon = \frac{A}{r}\Big(\frac{r}{D_I}\Big)^{d/2},\qquad \delta
= \Big(\frac{A_e}{A}\Big) < 1, \qquad 1 \geq Q \geq 0.
\end{equation}
Note that in $d=2$ space dimensions the noise amplitude $\epsilon$
is determined uniquely by the ratio of the replication rate $A$ to
replicator diffusion $D_I$. It is reasonable to assume that both
master sequence and error tail replicator molecules decay with the
\textsl{same} rate, $r = r_e$. Moreover, since both the master
sequence and error tail are built up from the same monomer pool
and have the same total length, they should also have identical
diffusion constants $D_I = D_e$. The much smaller monomers should
diffuse more rapidly, so we can take $D_s >> D_I=D_e \equiv D$.
Lastly, we note that the noise-averaged fields satisfy the
constraint:
\begin{equation}
\int d^2 \hat \mathbf{x} \langle \bar a + \bar b +
\frac{1}{\delta}\bar c \rangle = \bar N = \epsilon N,
\end{equation}
where $N$ is the total particle number.

The complete nondimensional model has five parameters: $\epsilon,
\delta, Q, D_s/D > 1$ and $\bar N$.

\end{document}